\documentclass[notitlepage]{revtex4-1}
\usepackage[utf8]{inputenc}
\usepackage[colorlinks=true,citecolor=blue,linkcolor=blue]{hyperref}
\usepackage[normalem]{ulem}
\usepackage{url}
\usepackage{graphicx,wrapfig,float,slashed,cancel}
\usepackage{amsmath,amssymb,epsfig,graphicx,xcolor,stmaryrd}
\usepackage{bm}
\usepackage{enumitem}
\usepackage{bbm}
\usepackage{multirow}
\definecolor{darkblue}{RGB}{1, 90, 173}

\begin{document}

%%%%%%%%%%%%%%%%%%%%%%%%%%% ---------- Title ---------- %%%%%%%%%%%%%%%%%%%%%%%%%%%

\title{Analysis of Fully Heavy $P_{(3c2b)}$ and $P_{(3b2c)}$ Pentaquark Candidates}

\author{K.~Azizi}
\thanks{Corresponding author}
\email{ E-mail: kazem.azizi@ut.ac.ir}
\affiliation{Department of Physics, University of Tehran, North Karegar Avenue, Tehran
14395-547, Iran}
\affiliation{Department of Physics, Dogus University, Dudullu-\"{U}mraniye, 34775
Istanbul, T\"{u}rkiye}
\affiliation{School of Particles and Accelerators, Institute for Research in Fundamental Sciences (IPM) P.O. Box 19395-5531, Tehran, Iran}
\author{Y.~Sarac}
\email{E-mail: yasemin.sarac@atilim.edu.tr}
\affiliation{Electrical and Electronics Engineering Department,
Atilim University, 06836 Ankara, T\"{u}rkiye}
\author{H.~Sundu}
\email{ E-mail: hayriyesundu.pamuk@medeniyet.edu.tr}
\affiliation{Department of Physics Engineering, Istanbul Medeniyet University, 34700 Istanbul, T\"{u}rkiye}

\date{\today}

\preprint{}

\begin{abstract}

Recent progress in experimental facilities, together with larger data samples and more refined analysis strategies has enabled the observation of many exotic hadronic states, adding new members to the hadron spectrum. Each newly reported signal encourages further experimental searches and simultaneously motivates theoretical studies aimed at uncovering additional nonconventional states.
Motivated by this perspective and by the increasing interest in systems containing multiple heavy quarks, we present a spectroscopic study of fully heavy pentaquark candidates with spin-parity quantum numbers $J^{P}=\frac{1}{2}^{-}$ and quark contents $QQQ'Q\bar{Q'}$, $QQQ'Q'\bar{Q}$, and $Q'Q'QQ\bar{Q}$, where $Q(Q')$ represents either $c(b)$ or $b(c)$ quarks. We employ the QCD sum rule approach with three different types of interpolating currents to obtain the corresponding masses and current coupling constants of the considered states. The following masses  for the states containing three $c$ and two $b$ quarks are predicted: $m_{(3c2b)}=14479.30\pm75.06~\mathrm{MeV}$ using the current $J_1$, $\tilde{m}_{(3c2b)}=14276.80\pm76.29~\mathrm{MeV}$ using $J_2$, and $\bar{m}_{(3c2b)}=14276.80\pm76.29~\mathrm{MeV}$ using $J_3$. The corresponding predictions for the states containing three $b$ and two $c$ quarks are  as $m_{(3b2c)}=17458.90\pm130.11~\mathrm{MeV}$ with $J_1$, $\tilde{m}_{(3b2c)}=17202.70\pm132.37~\mathrm{MeV}$ with $J_2$, and $\bar{m}_{(3b2c)}=17250.80\pm131.98~\mathrm{MeV}$ with $J_3$, respectively. In addition, we provide the corresponding current coupling constants, which can serve as useful inputs for analyses of decay properties and interaction mechanisms of these fully heavy pentaquark candidates.

\end{abstract}

%%%%%%%%%%%%%%%%%%%%%%%%%%%%%%%%%%%%%%%%%%%%%%%%%%%%%%%%%%%%%%%%%%%%%%%%%%%%%%%%%%

\maketitle

%\vspace{-1mm}
%%%%%%%%%%%%%%%%%%%%%%%%%%%%%%%%%
%%%%%%%%%%%%%%%%%%%%%%%%%%%%%%%%%
%\maketitle
\renewcommand{\thefootnote}{\#\arabic{footnote}}
\setcounter{footnote}{0}
%%%%%%%%%%%
\section{\label{sec:level1}Introduction}\label{intro}

The existence of exotic states with valence quark content different from that present in conventional mesons and baryons has been discussed since the early days of the quark model. Conventional mesons have valence quark structure as $q \bar{q}$, and baryons as three quarks (or three antiquarks). On the other hand, the theory of strong interaction does not restrict color singlets to only these minimal quark configurations. The possibility of the existence of such exotic states has kept the field active for decades. For a long time, however, the experimental situation was not decisive enough to establish a clear pattern about these states. A major milestone was reached  with the observation of the $X(3872)$ state in 2003~\cite{Choi:2003ue}. This observation provided a concrete starting point for a modern era of exotic hadron studies. In the following period, a growing number of exotic candidates were observed by different experiments~\cite{Aaij:2015tga,Aaij:2016ymb,Aaij:2019vzc,Zyla:2020zbs,LHCb:2020jpq,LHCb:2021vvq,LHCb:2021auc,LHCb:2021chn,LHCb:2022ogu}. Properties of these states have been compiled and updated in the PDG~\cite{ParticleDataGroup:2024cfk}. The interest over these exotic states has been increased due to these observations, which leads to extensive theoretical work aiming to clarify the internal structures and properties of these states which are still uncertain. The analyses attained in different pictures can often reproduce parts of the available data, so further careful analyses of masses, widths, quantum numbers, and decay patterns are needed to identify the properties of these states. Such studies provide information on the nature of already observed candidates, and help identify possible properties of new candidate states that are likely to be observed in the future as well. In other words, constructing solid information for prospective exotic states is essential for guiding upcoming experimental analyses. Additionally, understanding exotic hadrons provides insight into our understanding of confinement and the dynamics of the strong interaction.

LHCb collaboration reported the first observation of pentaquarks, which are among the members of the exotic state family, in 2015 through the amplitude analyses of $\Lambda_b^0 \rightarrow J/\psi p K^-$~\cite{Aaij:2015tga}.  These analyses resulted in identification of two states with masses and widths  $m_{P_c(4380)^+}=4380 \pm8 \pm 29~\mathrm{MeV}$, $\Gamma_{P_c(4380)^+}=205 \pm 18 \pm 86~\mathrm{MeV}$ and $m_{P_c(4450)^+}=4449.8 \pm 1.7 \pm 2.5~\mathrm{MeV}$, $\Gamma_{P_c(4450)^+}= 39 \pm 5 \pm 19~\mathrm{MeV}$~\cite{Aaij:2015tga}, respectively. These observations were supported by the full amplitude analyses in 2016~\cite{Aaij:2016ymb}. The 2019 update, based on a larger data sample, revealed an additional state, $P_c(4312)^+$, with
$m_{P_c(4312)^+}=4311.9 \pm 0.7^{ +6.8}_{-0.6}~\mathrm{MeV}$ and $\Gamma_{P_c(4312)^+}=9.8 \pm 2.7 ^{ +3.7}_{-4.5}~\mathrm{MeV}$~\cite{Aaij:2019vzc}. In the same analyses, the earlier state reported as $P_c(4450)$ was resolved into two nearby narrow peaks with masses and widths $m_{P_c(4440)^+}=4440.3 \pm 1.3^{+4.1}_{-4.7}~\mathrm{MeV}$, $\Gamma_{P_c(4440)^+}=20.6 \pm 4.9^{+8.7}_{-10.1}~\mathrm{MeV}$ and
$m_{P_c(4457)^+}=4457.3 \pm 0.6^{+4.1}_{-1.7}~\mathrm{MeV}$, $\Gamma_{P_c(4457)^+}=6.4 \pm 2.0^{+5.7}_{-1.9}~\mathrm{MeV}$.   These are followed by the evidence for a strange pentaquark candidate, $P_{cs}(4459)^0$, which was observed from the analyses of the $J/\psi\Lambda$ invariant mass spectrum in $\Xi_b^- \rightarrow J/\psi K^-\Lambda$ decays. This state has been reported to have mass $m=4458.8 \pm 2.9^{+4.7}_{-1.1}~\mathrm{MeV}$ and width $\Gamma=17.3 \pm 6.5^{+8.0}_{-5.7}~\mathrm{MeV}$~\cite{LHCb:2020jpq}. $B^{-}\rightarrow J/\psi \Lambda \bar{p}$ amplitude analyses revealed~\cite{LHCb:2022ogu} another member of pentaquark family reporting its measured mass and width as $4338.2 \pm 0.7\pm 0.4$~MeV and $ 7.0\pm 1.2 \pm 1.3$~MeV, respectively~\cite{LHCb:2022ogu}.

The discoveries of the pentaquark states have stimulated intense theoretical activities searching their spectrum and underlying structure. Investigations have been focused on shedding light on the arrangement of their constituents and the quantum numbers they carry. Enlightening the properties of these states serves also as guides for future experiments to uncover additional exotic hadrons. Moreover, given their valence quark content different from the conventional $qqq$ baryons and $q\bar{q}$ mesons, pentaquarks provide a valuable probe for the nonperturbative regime of QCD. These features of pentaquark states have driven many works, and they were investigated via different dynamical pictures and substructure assumptions. Among these structures assigned to pentaquark states is the diquark-diquark-antiquark~\cite{Lebed:2015tna,Maiani:2015vwa,Li:2015gta,Wang:2015ava,Anisovich:2015cia,Ghosh:2015ksa,Wang:2015ixb,Wang:2015epa,Wang:2015wsa,Wang:2016dzu,Zhang:2017mmw,Wang:2019got,Wang:2020eep,Ali:2020vee,Wang:2020rdh,Wang:2025qtm,Wang:2025fqh}. Their structure were also considered as diquark-triquark in Refs.~\cite{Wang:2016dzu,Zhu:2015bba}. Another commonly proposed structure was their being meson baryon molecules~\cite{Chen:2015loa,He:2015cea,Chen:2015moa,Roca:2015dva,Meissner:2015mza,Azizi:2016dhy,Wang:2018waa,Azizi:2018bdv,Liu:2019tjn,Wang:2019hyc,Chen:2020uif,Chen:2020kco,Chen:2020opr,Azizi:2020ogm,Peng:2020hql,Du:2021fmf,Wang:2021itn,Wang:2023eng,Li:2024wxr,Li:2024jlq,Liu:2024ugt,Hiyama:2025dfx,Clymton:2025hez}. Several analyses have argued that some observed structures might be explained as kinematical effects~\cite{Guo1,Guo2,Mikhasenko:2015vca,Liu1,Bayar:2016ftu,Co:2024szz}. Besides, applying the topological soliton model~\cite{Scoccola:2015nia} and a version of the D4-D8 model~\cite{Liu:2017xzo} their properties have been explored. In addition to the searches of the observed states, many studies have also proposed and analyzed additional pentaquark states that might be within reach of forthcoming experiments~\cite{Feijoo:2015kts,Chen:2015sxa,Chen:2016ryt,Lu:2016roh,Azizi:2017bgs,Rostami:2026jyz,Irie:2017qai,Azizi:2018dva,Cao:2019gqo,Gutsche:2019mkg,Zhang:2020vpz,Zhang:2020cdi,Liu:2020cmw,Paryev:2020jkp,Xie:2020ckr,Wang:2020bjt,Wang:2024yjp,Oset:2024fbk,Feijoo:2024qgq,Gordillo:2024blx,Yang:2024okq,Liu:2024dlc,Song:2024yli,Wang:2024brl,Liu:2025hvc,Wang:2025pjt,Ben:2025wqn,Liu:2025rqt,Roca:2025zyi,Chen:2025obf,Wang:2025hhx,Clymton:2025dzt,Guo:2020gkf,Jing:2025iqs,Clymton:2025zer,Liu:2025slt,Sharma:2025grd,Azizi:2025fmx,Li:2025hpd,Song:2025yut}. In Ref. \cite{Rostami:2026jyz}, comprehensive predictions on the properties of many possible pentaquark states regarding the flavor contents and spin-parity quantum numbers have been made using different machine learning techniques.

The discovery of additional exotic hadrons has become a natural expectation as experimental techniques and facilities continue to improve. New observations of such exotic states have further strengthened this expectation which has stimulated substantial theoretical activity. Taking different possible internal organization and quark content and using different frameworks such candidate states were explored~\cite{Rostami:2026jyz,Liu:2020cmw,Chen:2015sxa,Feijoo:2015kts,Lu:2016roh,Irie:2017qai,Chen:2016ryt,Zhang:2020cdi,Paryev:2020jkp,Gutsche:2019mkg,Azizi:2017bgs,Cao:2019gqo,Azizi:2018dva,Zhang:2020vpz,Wang:2020bjt,Xie:2020ckr,Duan:2024uuf,Kong:2024scz,Sharma:2024wpc,Zhang:2023teh,Yan:2023iie,Liu:2023clr,Wang:2023mdj,Wang:2023ael,Yang:2023dzb,Liu:2023oyc,Paryev:2023uhl,Lin:2023iww,Xin:2023gkf,Chen:2023qlx,Zhu:2023hyh,Yan:2023kqf,Sharma:2024pfi,Liu:2024mwn,Roca:2024nsi}. At the same time, the experimental observation of doubly heavy baryons and fully heavy exotic structures~\cite{LHCb:2017iph,LHCb:2018pcs,LHCb:2020bwg,Bouhova-Thacker:2022vnt,Zhang:2022toq,CMS:2023owd,ATLAS:2023bft} naturally encourages the idea that pentaquark resonances might also exist with multiple heavy valence quarks in their composition. This expectation provides strong motivation for investigating such systems in detail. In this study, we concentrate on fully heavy pentaquarks with quark substructures $QQQ'Q\bar{Q'}$, $QQQ'Q'\bar{Q}$, and $Q'Q'QQ\bar{Q}$, where $Q(Q')$ is either $c(b)$ or $b(c)$ quarks. Several studies have already explored the properties of fully heavy pentaquarks via various methods and structural assumptions~\cite{Zhang:2020vpz,Wang:2021xao,Yan:2021glh,An:2020jix,An:2022fvs,Zhang:2023hmg,Yang:2022bfu,Liang:2024met,Alonso-Valero:2024jim,Rashmi:2024ako,Gordillo:2024sem,Azizi:2024ito,Jing:2025iqs,Sharma:2025grd,Azizi:2025fmx,Jing:2025iqs,Azizi:2025fmx}. In Refs.~\cite{Zhang:2020vpz,Wang:2021xao,Azizi:2024ito} the fully charm or fully bottom pentaquark systems  were considered and their masses were predicted in the diquark diquark antiquark configuration as well as in the meson–baryon molecular picture, with different spin–parity assignments. The possible pentaquark states with spin parity quantum numbers $1/2^-$, $3/2^-$ or $5/2^-$ were investigated via the constituent quark model~\cite{Yan:2021glh,Zhang:2023hmg,Yang:2022bfu} to obtain their masses and bound state configurations. Application of the variational methods indicated no stable bound state~\cite{An:2022fvs}. With the chromomagnetic interaction analyses the spectra of the $cccc\bar{c}$ and $bbbb\bar{b}$ states were predicted~\cite{An:2020jix}. More recent investigations applying a nonrelativistic potential model~\cite{Liang:2024met}, a bound state Hartree Fock method~\cite{Alonso-Valero:2024jim}, an effective quark mass and screened charge framework~\cite{Rashmi:2024ako}, diffusion Monte Carlo calculations~\cite{Gordillo:2024sem} and extended G\"ursey–Radicati formalism~\cite{Rostami:2026jyz,Sharma:2025grd} have provided predictions on the masses and structures of these states.  Despite all these efforts, we need further analyses to clarify the nature of fully heavy pentaquark candidates and to provide theoretical insight that may guide future measurements. Such investigations also contribute to a better understanding of the nonperturbative regime of QCD and of heavy quark dynamics. In this work, we adopt the QCD sum rule method~\cite{Shifman:1978bx,Shifman:1978by,Ioffe81}, which takes place among the powerful nonperturbative frameworks. This method has shown strong performance across a wide range of hadronic applications providing predictions consistent with experimental observations. With this method we determine the masses and current coupling constants of fully heavy pentaquark states with spin parity $\frac{1}{2}^-$, composed of three $c(b)$ quarks and two $b(c)$ quarks, where one of these quarks will be an antiquark. With the chosen quark contents and the spin and parity quantum numbers, the interpolating currents are formed. For future experiments, the present results may offer quantitative input that helps shape such searches.

The paper is organized as follows. In Section~\ref{II}, we outline the QCD sum rule framework used to determine the masses and current coupling constants. Section~\ref{III} contains the numerical analysis and a discussion of the QCDSR results. The final section is devoted to the summary and conclusions.

\section{Sum rules for $QQQ'Q\bar{Q'}$, $QQQ'Q'\bar{Q}$ and $Q'Q'QQ\bar{Q}$ pentaquark states}\label{II}

In this section, we outline the main steps of the QCD sum rule analysis used to extract the masses and current coupling constants of fully heavy pentaquark states with quark contents $QQQ'Q\bar{Q'}$, $QQQ'Q'\bar{Q}$ and $Q'Q'QQ\bar{Q}$, where $Q(Q')$ stands for $c(b)$ or $b(c)$ heavy quarks. In what follows, these states are denoted by $P_{(3c2b)}$ and $P_{(3b2c)}$. To extract the masses and current couplings, we start from the following two point correlation function:
\begin{equation}
	\Pi(p)=i\int d^{4}xe^{ip\cdot
		x}\langle 0|\mathcal{T} \{J_{P_{(3Q2Q')}}(x)\bar{J}_{P_{(3Q2Q')}}(0)\}|0\rangle .
	\label{eq:CorrF1PQ}
\end{equation}
In the above expression, $\mathcal{T}$ denotes the time ordering operator, while $J_{P_{(3Q2Q')}}$ is the interpolating field for a fully heavy pentaquark built from three $Q$ quark and two $Q'$ quark fields, corresponding to the $P_{(3c2b)}$ or $P_{(3b2c)}$ states. In the present analyses following six interpolating currents in molecular form with different heavy quark configurations are adopted:
\begin{eqnarray}
	J_{1\,P_{(3Q2Q')}} &=& \left[\epsilon^{ijk}\,Q_i^{T}\,C\gamma_{\mu}\,Q_j\,\gamma_{5}\gamma^{\mu}\,Q'_k\right]\label{Eq:Current1}
	\left[\bar{Q}_{l}\, i\gamma_{5}\, Q'_{l}\right],\\
	J_{2\,P_{(3Q2Q')}} &=& \left[\epsilon^{ijk}\,Q_i^{T}\,C\gamma_{\mu}\,Q_j\,\gamma_{5}\gamma^{\mu}\,Q'_k\right]\label{Eq:Current2}
	\left[\bar{Q}'_{l}\, i\gamma_{5}\, Q_{l}\right],\\
	J_{3\,P_{(3Q2Q')}} &=& \left[\epsilon^{ijk}\,Q'{}_i^{T}\,C\gamma_{\mu}\,Q'_j\,\gamma_{5}\gamma^{\mu}\,Q_k\right]\label{Eq:Current3}
	\left[\bar{Q}_{l}\, i\gamma_{5}\, Q_{l}\right].
\end{eqnarray}
In Eqs.~(\ref{Eq:Current1}), (\ref{Eq:Current2}) and (\ref{Eq:Current3}), $T$ stands for the transpose operation, $i, j, k, l$ label the SU(3) color indices, and $C$ is the charge conjugation operator. The quark field $Q(Q')$ represents either $c(b)$ or $b(c)$ therefore these interpolating currents couple to pentaquark states with six different substructures. 

Within the QCD sum rule framework, the two point correlation function is analyzed in two parallel representations, namely hadronic and QCD sides. In hadronic side, it is  expressed through hadronic degrees of freedom, which provide mass and current coupling constant in the results. In the QCD side, the results are expressed through  the QCD coupling constant, quark masses, and quark-gluon condensates. After these calculations the results are matched via a dispersion relation. Considering the coefficients of same Lorentz structures on both sides, the QCD sum rules for the considered quantities are extracted. The results obtained from the calculations of the correlator contain information  about not only  the lowest state but also higher states and continuum, the effects of which are needed to be removed to get sole contribution from the lowest state. To achieve this aim, we use Borel transformation and continuum subtraction applications which provide improvement in the accuracy of results.

To calculate the hadronic side,  the interpolating currents are employed as an operator that couples to the hadronic state with the considered quantum numbers. In this side the correlator is represented in terms of the hadron mass and current coupling constant by inserting a complete set of hadronic states carrying the same quantum numbers as the interpolating current. This lead us to the following result in which we keep only the ground state contribution
\begin{eqnarray}
	\Pi^{\mathrm{Had}}_{n}(p)
	&=&
	\frac{
		\langle 0|J_{n\,P_{(3Q2Q')}}|P_{(3Q2Q')}(p,s)\rangle
		\langle P_{(3Q2Q')}(p,s)|\bar{J}_{n\,P_{(3Q2Q')}}|0\rangle
	}{
		m_{P_{(3Q2Q')}}^{2}-p^{2}
	}
	+\cdots,
	\label{eq:hadronic_side_P3Q2Qp}
\end{eqnarray}
where the dots stand for the contributions from higher resonances and the continuum and $n=1,~2,~3$ represent the interpolating currents given in Eqs.~(\ref{Eq:Current1}), (\ref{Eq:Current2}) and (\ref{Eq:Current3}), respectively. The ground-state pentaquark is represented with momentum $p$ and spin $s$ by $|P_{(3Q2Q')}(p,s)\rangle$ and the current coupling constant, $\lambda_n$, is introduced via the matrix element between the vacuum and the one-particle state as
\begin{eqnarray}
	\langle 0|J_{n\,P_{(3Q2Q')}}|P_{(3Q2Q')}(p,s)\rangle &=& \lambda_{n}\,u(p,s)\,.
	\label{eq:matrixelement_P3Q2Qp}
\end{eqnarray}
Using the parametrization of the vacuum to one particle matrix element, Eq.~(\ref{eq:matrixelement_P3Q2Qp}), in the  correlator, Eq.~(\ref{eq:hadronic_side_P3Q2Qp}), we next sum over the spin of the intermediate pentaquark state applying
\begin{eqnarray}\label{spin_sum}
	\sum_s u(p,s)\,\bar{u}(p,s) &=& \slashed{p}+ m_{P_{(3Q2Q')}} \,.
\end{eqnarray}
After these steps the correlator on the hadronic side becomes
\begin{eqnarray}\label{PhysSide_P3Q2Qp}
	\Pi^{\mathrm{Had}}_{n}(p)
	&=&
	\frac{\lambda_{n}^{2}\left(\slashed{p}+m_{P_{(3Q2Q')}}\right)}{p^{2}-m_{P_{(3Q2Q')}}^{2}}+\cdots \,.
\end{eqnarray}
A Borel transformation with respect to $-p^2$ turns the results into 
\begin{eqnarray}
	\mathcal{B}\Pi^{\mathrm{Had}}_{n}(p)
	&=&
	\lambda_{n}^{2}\,e^{-\frac{m_{P_{(3Q2Q')}}^{2}}{M^{2}}}
	\left(\slashed{p}+m_{P_{(3Q2Q')}}\right)+\cdots \,,
	\label{PhysSideF_P3Q2Qp}
\end{eqnarray}
with $M^2$ being the Borel parameter to be specified later from the numeric analyses.

On the QCD side, we evaluate the correlator by substituting the interpolating currents in Eqs.~(\ref{Eq:Current1}), (\ref{Eq:Current2}) and  (\ref{Eq:Current3})  directly into the correlation function. We apply Wick's theorem and carry out the contractions among quark fields. These lead us to the results:
\begin{eqnarray}
	\Pi_1^{\mathrm{QCD}}(p)
	&=&
	-i\int d^4x\,e^{ip\cdot x}\,
	\epsilon_{ijk}\,\epsilon_{i'j'k'}\,
	\Bigg\{
	2\,\mathrm{Tr}\!\Big[S_{Q'}^{ll'}(x)\gamma_{5}\,S_{Q}^{l'l}(-x)\gamma_{5}\Big]\,
	\mathrm{Tr}\!\Big[\gamma_{\nu}C\,S_{Q}^{T\,ii'}(x)\,C\gamma_{\mu}\,S_{Q}^{jj'}(x)\Big]\,
	\gamma_{5}\gamma^{\mu}\,S_{Q'}^{kk'}(x)\,\gamma^{\nu}\gamma_{5}
	\nonumber\\[2mm]
	&&\qquad
	-\,2\,\mathrm{Tr}\!\Big[\gamma_{\nu}C\,S_{Q}^{T\,ii'}(x)\,C\gamma_{\mu}\,S_{Q}^{jj'}(x)\Big]\,
	\gamma_{5}\gamma^{\mu}\,S_{Q'}^{kl'}(x)\,\gamma_{5}\,
	S_{Q}^{l'l}(-x)\,\gamma_{5}\,
	S_{Q'}^{l k'}(x)\,\gamma^{\nu}\gamma_{5}
	\Bigg\},
	\label{eq:PiQCD_two_terms}\label{Eq:QCD1}
\end{eqnarray}
\begin{eqnarray}
	\Pi_2^{\mathrm{QCD}}(p)
	&=&
	i\int d^4x\,e^{ip\cdot x}\,
	\epsilon_{ijk}\epsilon_{i'j'k'}\,
	\Big\{
	2\,\mathrm{Tr}\!\left[
	S_{Q}^{ii'}(x)\,\gamma_{\nu}C\,
	S_{Q}^{T\,jj'}(x)\,C\gamma_{\mu}
	\right]\,
	\mathrm{Tr}\!\left[
	i\gamma_{5}\,S_{Q}^{ll'}(x)\,
	i\gamma_{5}\,S_{Q'}^{l'l}(-x)
	\right]\,
	\gamma_{5}\gamma^{\mu}\,S_{Q'}^{kk'}(x)\,\gamma_{5}\gamma^{\nu}
	\nonumber\\[2mm]
	&&
	-\,2\,\mathrm{Tr}\!\left[
	\gamma_{\nu}C\,S_{Q}^{T\,ii'}(x)\,C\gamma_{\mu}\,S_{Q}^{jl'}(x)\,
	i\gamma_{5}\,S_{Q'}^{l'l}(-x)\,
	i\gamma_{5}\,S_{Q}^{l j'}(x)
	\right]\,
	\gamma_{5}\gamma^{\mu}\,S_{Q'}^{kk'}(x)\,\gamma_{5}\gamma^{\nu}
	\nonumber\\[2mm]
	&&
	-\,2\,\mathrm{Tr}\!\left[
	i\gamma_{5}\,S_{Q}^{l i'}(x)\,
	\gamma_{\nu}C\,S_{Q}^{T\,jj'}(x)\,C\gamma_{\mu}\,S_{Q}^{il'}(x)\,
	i\gamma_{5}\,S_{Q'}^{l'l}(-x)
	\right]\,
	\gamma_{5}\gamma^{\mu}\,S_{Q'}^{kk'}(x)\,\gamma_{5}\gamma^{\nu}
	\Big\},
	\label{eq:PiQCD_three_terms_new}\label{Eq:QCD2}
\end{eqnarray}
\begin{eqnarray}
	\Pi_3^{\mathrm{QCD}}(p)
	&=&
	i\int d^4x\,e^{ip\cdot x}\,
	\epsilon_{ijk}\epsilon_{i'j'k'}\,
	\Big\{
	2\,\mathrm{Tr}\!\left[
	S_{Q'}^{jj'}(x)\,\gamma_{\nu}C\,
	S_{Q'}^{T\,ii'}(x)\,C\gamma_{\mu}
	\right]\,
	\mathrm{Tr}\!\left[
	S_{Q}^{ll'}(x)\,i\gamma_{5}\,
	S_{Q}^{l'l}(-x)\,i\gamma_{5}
	\right]\,
	\gamma_{5}\gamma^{\mu}\,S_{Q}^{kk'}(x)\,\gamma_{5}\gamma^{\nu}
	\nonumber\\[2mm]
	&&
	-\,2\,\mathrm{Tr}\!\left[
	S_{Q'}^{ij'}(x)\,\gamma_{\nu}C\,
	S_{Q'}^{T\,ji'}(x)\,C\gamma_{\mu}
	\right]\,
	\gamma_{5}\gamma^{\mu}\,S_{Q}^{kl'}(x)\,
	i\gamma_{5}\,S_{Q}^{l'l}(-x)\,i\gamma_{5}\,S_{Q}^{l k'}(x)\,
	\gamma_{5}\gamma^{\nu}
	\Big\}.
	\label{eq:PiQCD_thirdterm_sameform}\label{Eq:QCD3}
\end{eqnarray}
In these equations $\Pi_n^{\mathrm{QCD}}(p)$ with $n=1,~2,~3$ is the result of the contractions obtained from the interpolarting currents given in Eqs.~(\ref{Eq:Current1}), (\ref{Eq:Current2}) and  (\ref{Eq:Current3}), respectively and $S_{Q(Q')}(x)$ is the heavy quark propagator used for $c$ or $b$ quarks according to the quark structure of the interpolating current applied in the calculation. The heavy quark propagator is given as
\begin{eqnarray}
	S_{Q}^{ab}(x)&=&\frac{i}{(2\pi)^4}\int d^4k e^{-ik \cdot x} \left\{
	\frac{\delta_{ab}}{\!\not\!{k}-m_Q}
	-\frac{g_sG^{\alpha\beta}_{ab}}{4(k^2-m_Q^2)^2}[\sigma_{\alpha\beta}(\!\not\!{k}+m_Q)+
	(\!\not\!{k}+m_Q)\sigma_{\alpha\beta}]\right.\nonumber\\
	&&\left.+\frac{\pi^2}{3} \langle \frac{\alpha_sGG}{\pi}\rangle
	\delta_{ab}m_Q \frac{k^2+m_Q\!\not\!{k}}{(k^2-m_Q^2)^4}
	+\cdots\right\}.
	\label{eq:Qpropagator}
\end{eqnarray}
In Eq.~(\ref{eq:Qpropagator}), we use the standard notations $G_{ab}^{\alpha\beta}=G_A^{\alpha\beta}t_{ab}^A$, $GG\equiv G^A_{\mu\nu}G^{A,\mu\nu}$, and $t^A=\lambda^A/2$, where $\lambda^A$ are the Gell-Mann matrices, with $a,~b=1,~2,~3$ and $A=1,~2,\cdots,~8$. Insertion of the propagator into Eqs.~(\ref{Eq:QCD1}), (\ref{Eq:QCD2}) and (\ref{Eq:QCD3}) is followed by applications of the Fourier and  Borel transformations. In the last step, the continuum is removed under the quark hadron duality assumption. Because these procedures lead us long expressions, we prefer to give numerical analyses of the results in  next section instead of giving their overwhelmingly long expressions. The correlator naturally splits into two Lorentz structures $\slashed{p}$ and $\mathbbm{1}$. Both are applicable in extraction of the masses and  corresponding current couplings of the considered states. In our analyses we consider the Lorentz structure $\slashed{p}$ and represent the results of calculations for this structure as: 
\begin{eqnarray}
	\mathcal{B}\Pi^{\mathrm{QCD}}_{n}(s_0,M^2)=\int_{(3m_Q+2m_{Q'})^2}^{s_0} ds e^{-\frac{s}{M^2}}\rho_{n}(s)\, , 
	\label{Eq:Cor:QCD}
\end{eqnarray}
where $\rho_{n}(s)$ denotes the spectral density obtained by taking the imaginary part of the coefficients of $\slashed{p}$ structures in the QCD side of the results calculated from  Eqs.~(\ref{Eq:QCD1}), (\ref{Eq:QCD2}) and (\ref{Eq:QCD3}) and given as $\rho_{n}(s)=\frac{1}{\pi}\mathrm{Im}\left[\Pi_{n}^{\mathrm{QCD}}\right]$ with  $n=1,~2,~3$ .

By matching the coefficients of the same Lorentz structures, $\slashed{p}$, in the QCD and hadronic representations, we obtain the QCD sum rules for the mass and the current coupling constant of the states considered here as: 
\begin{eqnarray}
	\lambda_{n}^{2}\,e^{-\frac{m_{P_{(3Q2Q')}}^{2}}{M^{2}}}
	=
	\mathcal{B}\Pi^{\mathrm{QCD}}_{n}(s_{0},M^{2}),
	\label{QCDsumrule_P3Q2Qp}
\end{eqnarray}
where $\mathcal{B}\Pi^{\mathrm{QCD}}_{n}(s_{0},M^{2})$ denotes the Borel transformed and continuum subtracted QCD sides corresponding to the chosen Lorentz structure. Using these relations the masses and  current coupling constants are obtained from
\begin{eqnarray}
	m_{P_{(3Q2Q')}}^{2}
	=
	\frac{
		\frac{d}{d\left(-\frac{1}{M^{2}}\right)}
		\mathcal{B}\Pi^{\mathrm{QCD}}_{n}(s_{0},M^{2})
	}{
		\mathcal{B}\Pi^{\mathrm{QCD}}_{n}(s_{0},M^{2})
	},
	\label{mass_P3Q2Qp}
\end{eqnarray}
and 
\begin{eqnarray}
	\lambda_{n}^{2}
	=
	e^{\frac{m_{P_{(3Q2Q')}}^{2}}{M^{2}}}\,
	\mathcal{B}\Pi^{\mathrm{QCD}}_{n}(s_{0},M^{2}),
	\label{lambda_P3Q2Qp}
\end{eqnarray}
respectively.

At this point, we have all the necessary expressions, so the next stage is getting numerical results of the physical quantities of interest which will be presented in the next section.

\section{Numeric evaluation of sum rules}\label{III}

\begin{figure} []
	\centering
	\begin{tabular}{cccc}
		\includegraphics[totalheight=5cm,width=7cm]{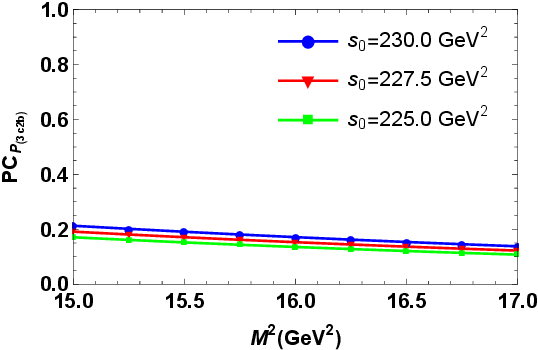} &
	    \includegraphics[totalheight=5cm,width=7cm]{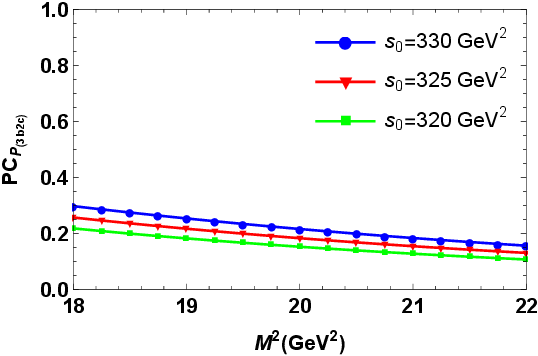} \\
	    \textbf{(a)}  & \textbf{(b)} 
	\end{tabular}
	\begin{tabular}{cccc}
		\includegraphics[totalheight=5cm,width=7cm]{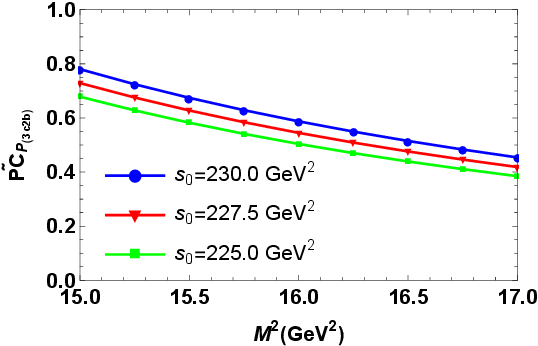} &
		\includegraphics[totalheight=5cm,width=7cm]{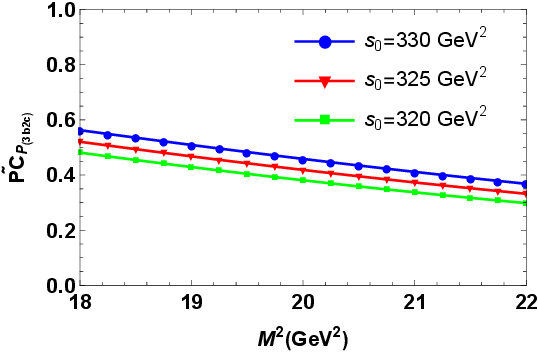}   \\
		\textbf{(c)}  & \textbf{(d)}  \\[6pt]
	\end{tabular}
		\begin{tabular}{cccc}
		\includegraphics[totalheight=5cm,width=7cm]{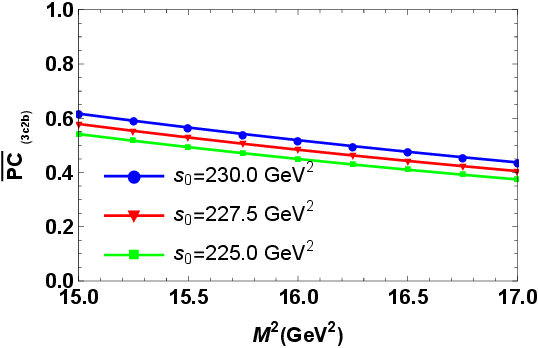} &
		\includegraphics[totalheight=5cm,width=7cm]{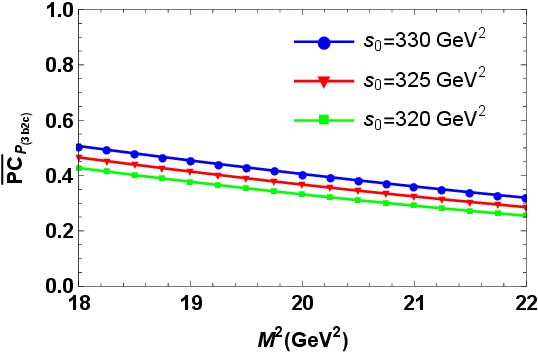}   \\
		\textbf{(e)}  & \textbf{(f)}  \\[6pt]
	\end{tabular}
	\caption{ The PC results obtained as a function of the Borel parameter, $M^2$, at different values of threshold parameters, $s_0$, for  
		\textbf{(a)} $P_{(3c2b)}$ and
		\textbf{(b)} $P_{(3b2c)}$ from current 1, 
		\textbf{(c)} $P_{(3b2c)}$ and
		\textbf{(d)} $P_{(3c2b)}$ from current 2,
		\textbf{(e)} $P_{(3c2b)}$ and
		\textbf{(f)} $P_{(3b2c)}$ from current 3.
		}
	\label{fig:1}
\end{figure}

\begin{figure} []
	\centering
	\begin{tabular}{cccc}
		\includegraphics[totalheight=5cm,width=7cm]{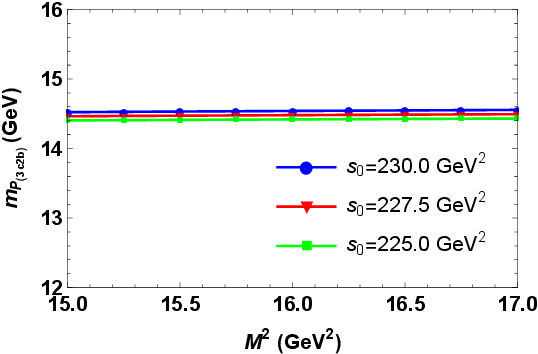} &
		\includegraphics[totalheight=5cm,width=7cm]{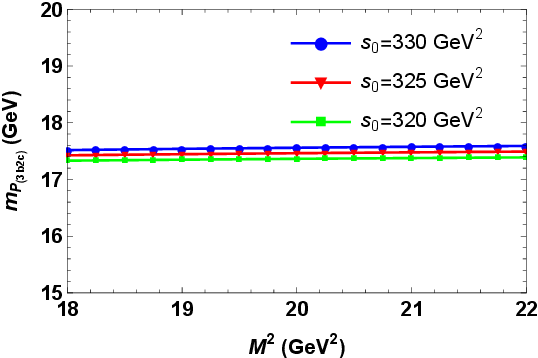} \\
		\textbf{(a)}  & \textbf{(b)}  \\[6pt]
	\end{tabular}
	\begin{tabular}{cccc}
		\includegraphics[totalheight=5cm,width=7cm]{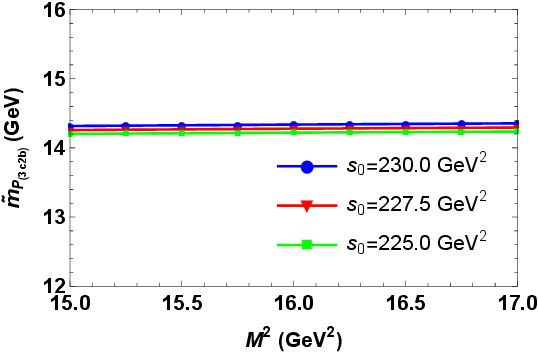} &
		\includegraphics[totalheight=5cm,width=7cm]{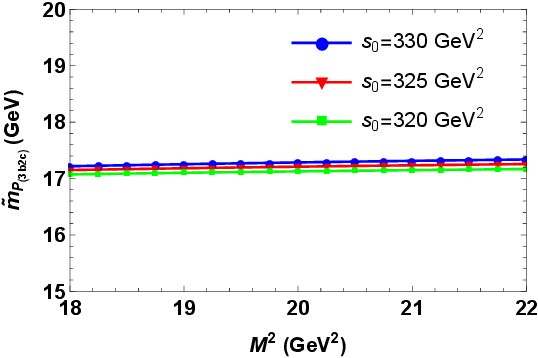} \\
		\textbf{(c)}  & \textbf{(d)}  \\[6pt]
	\end{tabular}
	\begin{tabular}{cccc}
		\includegraphics[totalheight=5cm,width=7cm]{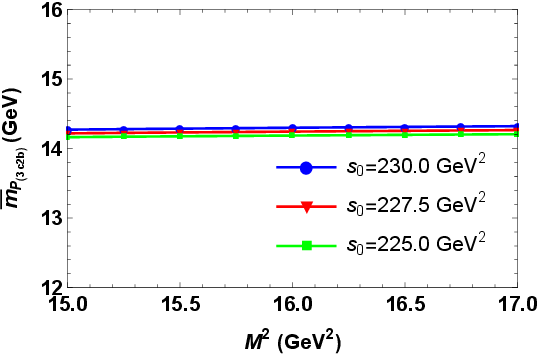} &
		\includegraphics[totalheight=5cm,width=7cm]{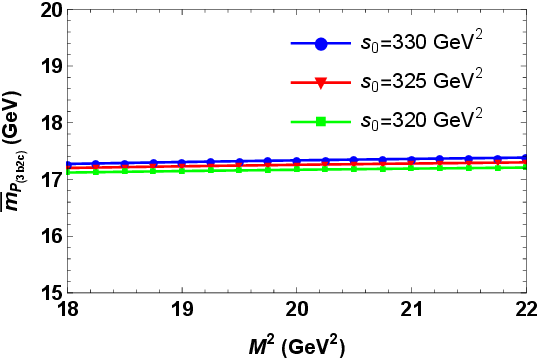} \\
		\textbf{(e)}  & \textbf{(f)}  \\[6pt]
	\end{tabular}
	\caption{ The masses of the considered pentaquark states as a function of the Borel parameter $M^2$ for various $s_0$ values obtained for
		(a) $P_{(3c2b)}$ and
		(b) $P_{(3b2c)}$ states using current 1,
		(c) $P_{(3c2b)}$ and
		(d) $P_{(3b2c)}$ states using current 2, 
		(e) $P_{(3c2b)}$ and
		(f) $P_{(3b2c)}$ states using current 3.
	}
	\label{fig:massvsMsq}
\end{figure}

\begin{figure} []
	\centering
	\begin{tabular}{cccc}
		\includegraphics[totalheight=5cm,width=7cm]{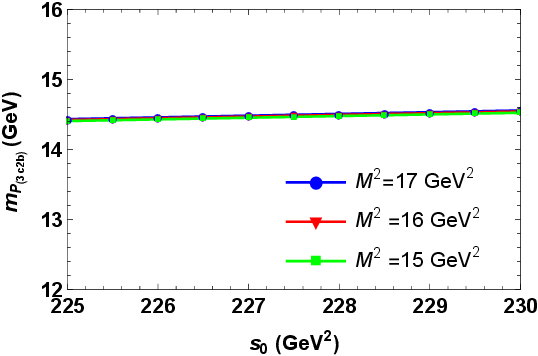} &
		\includegraphics[totalheight=5cm,width=7cm]{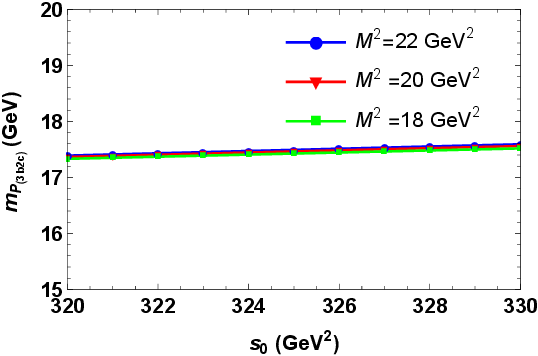} \\
		\textbf{(a)}  & \textbf{(b)}  \\[6pt]
	\end{tabular}
	\begin{tabular}{cccc}
		\includegraphics[totalheight=5cm,width=7cm]{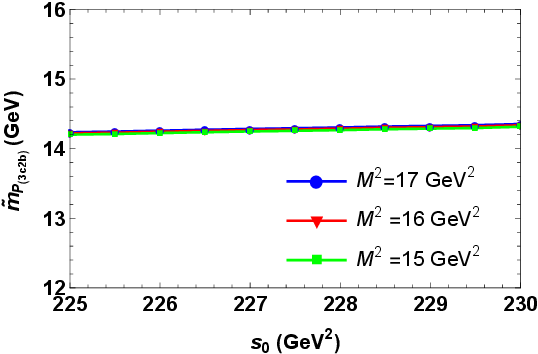} &
		\includegraphics[totalheight=5cm,width=7cm]{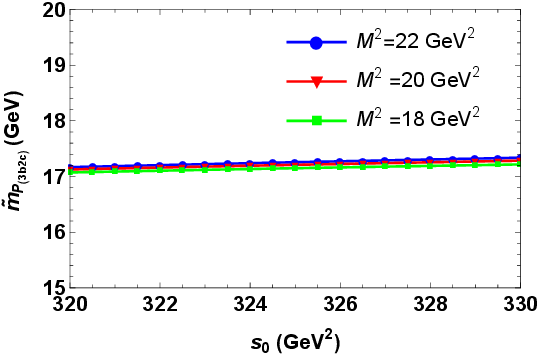} \\
		\textbf{(c)}  & \textbf{(d)}  \\[6pt]
	\end{tabular}
	\begin{tabular}{cccc}
		\includegraphics[totalheight=5cm,width=7cm]{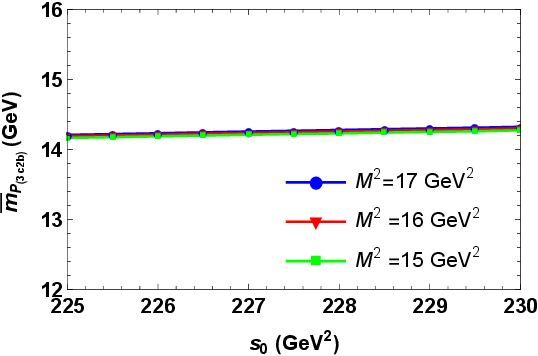} &
		\includegraphics[totalheight=5cm,width=7cm]{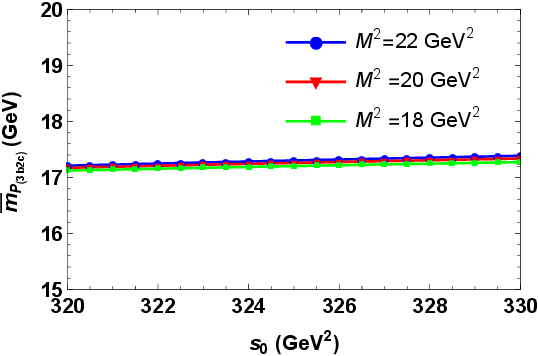} \\
		\textbf{(e)}  & \textbf{(f)}  \\[6pt]
	\end{tabular}
	\caption{ The masses of the considered pentaquark states as a function of the threshold parameter, $s_0$, for various $M^2$ values obtained for
		(a) $P_{(3c2b)}$ and
		(b) $P_{(3b2c)}$ states using current 1,
		(c) $P_{(3c2b)}$ and
		(d) $P_{(3b2c)}$ states using current 2, 
		(e) $P_{(3c2b)}$ and
		(f) $P_{(3b2c)}$ states using current 3.
	}
	\label{fig:massvss0}
\end{figure}

\begin{figure} []
	\centering
	\begin{tabular}{cccc}
		\includegraphics[totalheight=5cm,width=7cm]{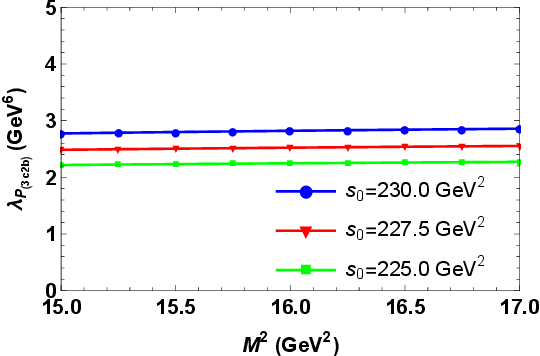} &
		\includegraphics[totalheight=5cm,width=7cm]{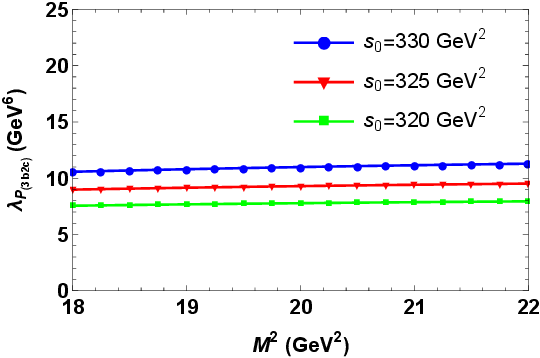} \\
		\textbf{(a)}  & \textbf{(b)}  \\[6pt]
	\end{tabular}
	\begin{tabular}{cccc}
		\includegraphics[totalheight=5cm,width=7cm]{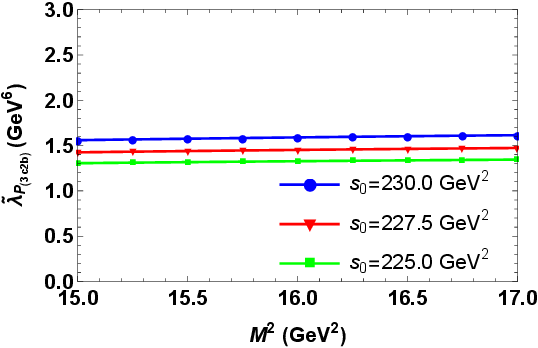} &
		\includegraphics[totalheight=5cm,width=7cm]{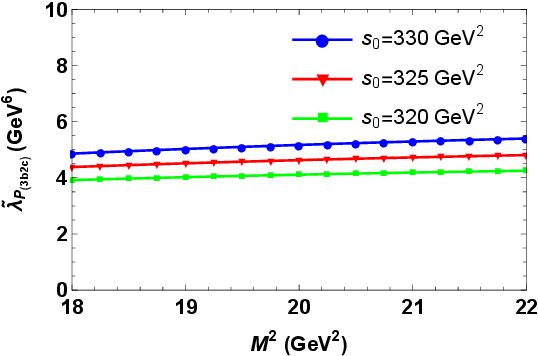} \\
		\textbf{(c)}  & \textbf{(d)}  \\[6pt]
	\end{tabular}
	\begin{tabular}{cccc}
		\includegraphics[totalheight=5cm,width=7cm]{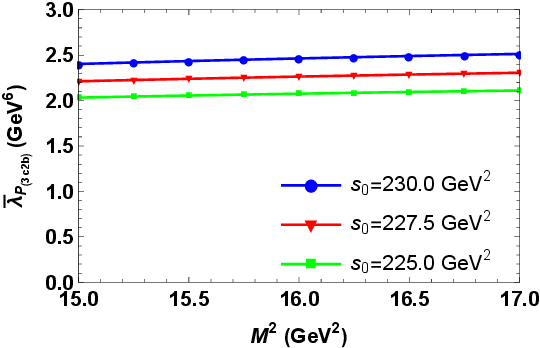} &
		\includegraphics[totalheight=5cm,width=7cm]{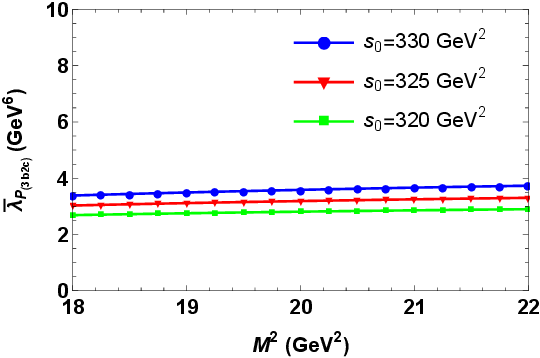} \\
		\textbf{(e)}  & \textbf{(f)}  \\[6pt]
	\end{tabular}
	\caption{ 
		 The current coupling constants  of the considered pentaquark states as a function of the Borel parameter, $M^2$, for various $s_0$ values obtained for
		(a) $P_{(3c2b)}$ and
		(b) $P_{(3b2c)}$ states using current 1,
		(c) $P_{(3c2b)}$ and
		(d) $P_{(3b2c)}$ states using current 2, 
		(e) $P_{(3c2b)}$ and
		(f) $P_{(3b2c)}$ states using current 3.
	}
	\label{fig:resvsMsq}
\end{figure}

\begin{figure} []
	\centering
	\begin{tabular}{cccc}
		\includegraphics[totalheight=5cm,width=7cm]{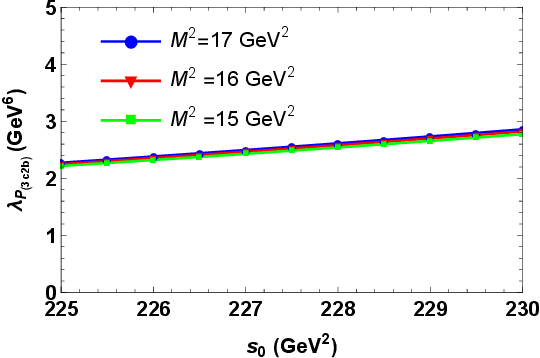} &
		\includegraphics[totalheight=5cm,width=7cm]{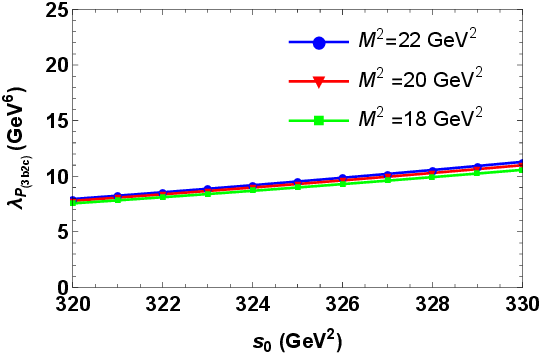} \\
		\textbf{(a)}  & \textbf{(b)}  \\[6pt]
	\end{tabular}
	\begin{tabular}{cccc}
		\includegraphics[totalheight=5cm,width=7cm]{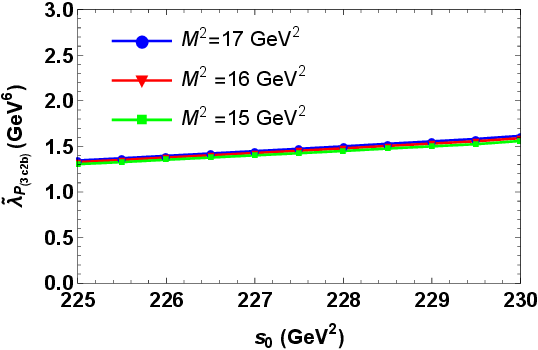} &
		\includegraphics[totalheight=5cm,width=7cm]{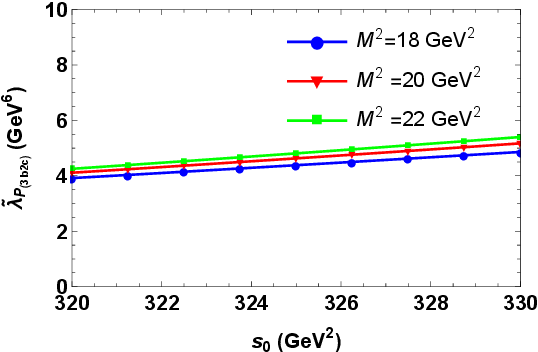} \\
		\textbf{(c)}  & \textbf{(d)}  \\[6pt]
	\end{tabular}
	\begin{tabular}{cccc}
		\includegraphics[totalheight=5cm,width=7cm]{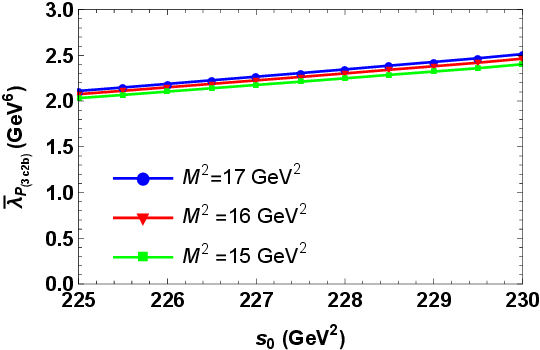} &
		\includegraphics[totalheight=5cm,width=7cm]{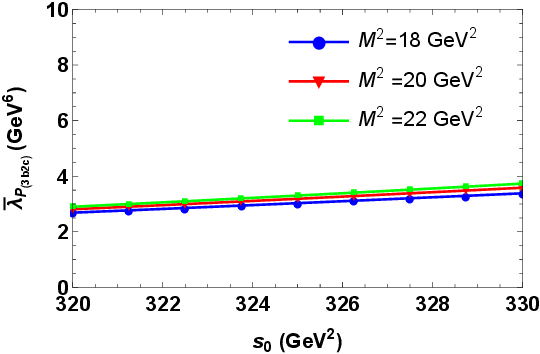} \\
		\textbf{(e)}  & \textbf{(f)}  \\[6pt]
	\end{tabular}
	\caption{ 
		The current coupling constants of the considered pentaquark states as a function of threshold parameter, $s_0$,  for various $M^2$ values obtained for 
		(a) $P_{(3c2b)}$ and
		(b) $P_{(3b2c)}$ states using current 1,
		(c) $P_{(3c2b)}$ and
		(d) $P_{(3b2c)}$ states using current 2, 
		(e) $P_{(3c2b)}$ and
		(f) $P_{(3b2c)}$ states using current 3.
	}
	\label{fig:resvss0}
\end{figure}

In this section, we use the results obtained in previous section together with the necessary input parameters to get their numerical values. Some of these input parameters are listed in Table~\ref{tab:Inputs}.
\begin{table}[h!]
	%\rowcolors{1}{lightgray}{white}
	\begin{tabular}{|c|c|}
		\hline\hline
		Parameters & Values \\ \hline\hline
		$m_{c}$                                    & $1.2730\pm 0.0046~\mathrm{GeV}$ \cite{ParticleDataGroup:2024cfk}\\
		$m_{b}$                                     & $4.18^{+0.04}_{-0.03}~\mathrm{GeV}$ \cite{ParticleDataGroup:2024cfk}\\
		$\langle \frac{\alpha_s}{\pi} G^2 \rangle $ & $(0.012\pm0.004)$ $~\mathrm{GeV}^4 $\cite{Belyaev:1982cd}\\
		%$\langle g_s^3 G^3 \rangle $               & $ (0.57\pm0.29)$ $~\mathrm{GeV}^6 $\cite{Narison:2015nxh}\\
		\hline\hline
	\end{tabular}%
	\caption{The set of input parameters required for the numerical evaluation.}
	\label{tab:Inputs}
\end{table} 
Nevertheless, this input set is not sufficient by itself, since two additional auxiliary parameters, namely threshold  and Borel parameters, must also be determined. The threshold parameter $s_0$ enters into the calculation  through the quark hadron duality ansatz, and the Borel parameter $M^2$ originates from the Borel transformation. Their working regions are determined by imposing the usual criteria of the QCD sum rules method. To choose the $s_0$, being closely connected to energy of excited states, our guide is the  expected location of the first possible excited state. Therefore, we select $s_0$ by estimating the energies of the possible first excited pentaquark states and based on these considerations, we work with the following interval:
\begin{eqnarray}
225~\mathrm{GeV}^2  \leq s_0  \leq 230~\mathrm{GeV}^2, 
\end{eqnarray}
for the states including three $c$ and two $b$ quarks, and
\begin{eqnarray}
320~\mathrm{GeV}^2 \leq s_0  \leq 330~\mathrm{GeV}^2, 
\end{eqnarray}
for that  including three $b$ and two $c$ quarks. The Borel parameter interval is fixed by imposing the usual constraints of the method: OPE convergence, pole dominance, and suppression of the continuum and higher resonances. The lower bound of the Borel parameter is fixed by requiring that the higher dimensional condensate contributions do not dominate the sum rule.  The upper bound is determined by requiring the pole contribution to dominate over the continuum and higher excited states.
To this end, we use the pole contribution relation given as:
\begin{equation}
	PC=
	\frac{\mathcal{B} \Pi^{\mathrm{QCD}}(s_0,M^2)}
	{\mathcal{B} \Pi^{\mathrm{QCD}}(\infty,M^2)}.
\end{equation}
The upper bound of $M^{2}$ is then set such that $PC$ stays as high as possible in the working interval. For our cases we fix upper bound by taking $PC \gtrsim 12\%$ for all states, which is commonly regarded as acceptable for pentaquark analyses. To show our $PC$ results, we provide Fig.~\ref{fig:1}.
Finally, to get  reliable results, the masses and residues extracted from the sum rules should not exhibit a strong sensitivity to the chosen Borel and threshold parameters inside  their working regions. Our results satisfy this requirement  within the intervals of the Borel parameters:
\begin{eqnarray}
	15.0~\mbox{GeV}^2\leq M^2\leq 17.0~\mbox{GeV}^2,
\end{eqnarray}
for $P_{(3c2b)}$ pentaquark states and
\begin{eqnarray}
	18.0~\mbox{GeV}^2\leq M^2\leq 22.0~\mbox{GeV}^2,
\end{eqnarray}
for $P_{(3b2c)}$ pentaquark states, respectively. To check how sensitive our predictions are to these auxiliary parameters, we plot the extracted quantities versus Borel parameter and the continuum threshold graphs which are presented in Figs.~\ref{fig:massvsMsq}, \ref{fig:massvss0}, \ref{fig:resvsMsq} and \ref{fig:resvss0}.

After establishing the proper ranges of the auxiliary parameters, we use them to extract the numerical predictions for the masses and current coupling constants, including their errors. The uncertainties of the input parameters, together with those coming from the auxiliary parameter intervals, constitute the main sources of the errors in our results. Our findings are summarized in Table~\ref{tab:table2}.
\begin{table}[]
	\centering
	\begin{tabular}{|c|c|c|c|}
		\hline
		Current & State & Mass~(MeV) & $\lambda~(\mathrm{GeV}^6)$ \\ \hline \hline
		\multirow{2}{*}{$J_1$}
		& $P_{(3c2b)}$ & $14479.30 \pm 75.06$  & $2.53 \pm 0.32$  \\ \cline{2-4}
		& $P_{(3b2c)}$ & $17458.90 \pm 130.11$ & $9.33 \pm 1.87$  \\ \hline \hline
		
		\multirow{2}{*}{$J_2$}
		& $P_{(3c2b)}$ & $14276.80 \pm 76.29$  & $1.46 \pm 0.16$  \\ \cline{2-4}
		& $P_{(3b2c)}$ & $17202.70 \pm 132.37$ & $4.62 \pm 0.74$  \\ \hline \hline
		
		\multirow{2}{*}{$J_3$}
		& $P_{(3c2b)}$ & $14276.80 \pm 76.29$  & $2.27 \pm 0.24$  \\ \cline{2-4}
		& $P_{(3b2c)}$ & $17250.80 \pm 131.98$ & $3.18 \pm 0.52$  \\ \hline
	\end{tabular}
	\caption{Masses and current coupling constants obtained from the QCD sum rules for the considered fully heavy pentaquark states using the currents $J_1$, $J_2$, and $J_3$.}
	\label{tab:table2}
\end{table}

In various studies, the pentaquark states with similar quark contents and quantum numbers were investigated which are presented in Table~\ref{tab:mass-comp-bycurrents}. 
\begin{table*}[t]
	\centering
	\renewcommand{\arraystretch}{1.15}
	\begin{tabular}{|c|c|c|c|c|c|}
		\hline
		Quark content & This work
		& Ref.~\cite{Gordillo:2024blx}
		& Ref.~\cite{An:2020jix}
		& Ref.~\cite{An:2022fvs}
		& Ref.~\cite{Zhang:2023hmg} \\
		\hline \hline
		
		\multicolumn{1}{|c|}{} & \multicolumn{1}{|c|}{$J_1$}
		& \multicolumn{1}{|c|}{} & \multicolumn{1}{|c|}{} & \multicolumn{1}{|c|}{} & \multicolumn{1}{|c|}{} \\
		\hline
		$ccbb\bar{c}$ & $14479.30\pm75.06$
		& 14348
		& 14406, 14318, 14253, 14185
		& -
		& 14893, 14875, 14852, 14821 \\
		\hline
		$bbcc\bar{b}$ & $17458.90\pm130.11$
		& 17536
		& 17576, 17496, 17437, 17405
		& -
		& 18200, 18180, 18168, 18154 \\
		\hline \hline
		
		\multicolumn{1}{|c|}{} & \multicolumn{1}{|c|}{$J_2$}
		& \multicolumn{1}{|c|}{} & \multicolumn{1}{|c|}{} & \multicolumn{1}{|c|}{} & \multicolumn{1}{|c|}{} \\
		\hline
		$ccbc\bar{b}$ & $14276.80\pm76.29$
		& 14206
		& 14411, 14357, 14238
		& -
		& 14895, 14878, 14862 \\
		\hline
		$bbcb\bar{c}$ & $17202.70\pm132.37$
		& 17448
		& 17578, 17523, 17399
		& -
		& 18198, 18177, 18159 \\
		\hline \hline
		
		\multicolumn{1}{|c|}{} & \multicolumn{1}{|c|}{$J_3$}
		& \multicolumn{1}{|c|}{} & \multicolumn{1}{|c|}{} & \multicolumn{1}{|c|}{} & \multicolumn{1}{|c|}{} \\
		\hline
		$bbcc\bar{c}$ & $14276.80\pm76.29$
		& 14248
		& 14406, 14318, 14253, 14185
		& 14566.0
		& - \\
		\hline
		$ccbb\bar{b}$ & $17250.80\pm131.98$
		& 17486
		& 17576, 17496, 17437, 17405
		& 17784.5
		& - \\
		\hline
	\end{tabular}
	\caption{Comparison of the $J^{P}=\frac{1}{2}^{-}$ fully heavy pentaquark masses in unit of MeV arranged according to the currents $J_{1}$, $J_{2}$, and $J_{3}$. }
	\label{tab:mass-comp-bycurrents}
\end{table*}
From Table~\ref{tab:mass-comp-bycurrents}, one sees that our $J^{P}=\frac{1}{2}^{-}$ masses are lower than the corresponding values in Refs.~\cite{Zhang:2023hmg,An:2022fvs}. Our predictions lie close to the lower edge of the reported results and can be particularly near the smallest quoted values in the $ccbc\bar b$ case given in Refs.~\cite{An:2020jix,Gordillo:2024blx}.

\section{Discussion and Concluding remarks}\label{IV}

The spectrum of exotic hadrons continues to expand as a result of the steady improvement of experimental facilities, larger data sets, and refined analysis techniques. Each new observation motivates the theoretical investigations of new quark configurations that can guide upcoming searches. In this work, by the growing interest in systems containing heavy quarks, we investigated fully heavy pentaquark candidates with valence contents $ccbb\bar c$, $bbcc\bar b$,  $ccbc\bar b$, $bbcb\bar c$, $bbcc\bar c$, $ccbb\bar c$,  and spin parity $J^{P}=\frac{1}{2}^{-}$, denoted as $P_{(3c2b)}$ and $P_{(3b2c)}$, respectively. Using the QCD sum rules method, the masses of these candidate pentaquark states were calculated applying three different interpolating currents. 

Considering the mentioned quark contents, for three different interpolating currents, we  obtained the mass values as follows: $m_{(3c2b)}=14479.30\pm75.06~\text{MeV}$ and $m_{(3b2c)}=17458.90\pm130.11~\text{MeV}$ using the current $J_1$, $\tilde{m}_{(3c2b)}=14276.80\pm76.29~\text{MeV}$ and $\tilde{m}_{(3b2c)}=17202.70\pm132.37~\text{MeV}$ using the current $J_2$, and $\bar{m}_{(3c2b)}=14276.80\pm76.29~\text{MeV}$ and $\bar{m}_{(3b2c)}=17250.80\pm131.98~\text{MeV}$ using the current $J_3$.

To gain a more complete picture of such states, further studies are clearly needed. To understand their internal structure and properties and to provide guidance for forthcoming experiments, it is necessary to analyze these states within various theoretical frameworks. In this respect, the results presented here may serve for comparisons with forthcoming studies and provide useful input for future investigations of the interaction mechanisms of these states.

%%%%%%%%%%%%%%%%%%%%%%%%%%%%

\section*{ACKNOWLEDGMENTS} 
K. Azizi thanks Iran national science foundation (INSF) for the partial financial support supplied under the elites Grant No. 40405095.

%%%%%%%%%%%%%%%%%%%%%%%%%%%%%%

%%%%%%%%%%%%%%%%%%%%%%%%%%%%%%
\newpage
%\vspace{2mm} {\it Acknowledgments.}--- {\small .}
\bibliographystyle{apsrev4-1}  % TeX Live 2015 ile uyumlu
\bibliography{refs3Q2Qp1a}
%%%%%%%%%%%%%%%%%%%%%%%%%%%%%%%%%%%%%%%%%%%%%%%%%%%%%%%%%%%%%%%%%%%%%%%%%%%%%%%%%%

%%%%%%%%%%%%%%%%%%%%%%%%%%%%%%%%%%%%%%%%%%%%%%%%%%%%%%%%%%%%%%%%%%%%%%%%%%%%%%%%%%

\end{document}